\pdfoutput=1

\documentclass[a4paper,11pt]{article}
\usepackage{hyperref}
\usepackage{graphicx,subfigure}
\usepackage{amsmath,amssymb,dsfont}

\usepackage{algorithmic,algorithm}

\floatname{algorithm}{Procedure}

\algsetup{indent=1em}

\usepackage{listings}
\usepackage{graphicx}
\newcommand{\Mathematica}{\emph{Mathematica}}

\newcommand{\TRQS}{\InlineCode{TRQS}}
\newcommand{\InlineCode}[1]{{\small \textbf{#1}}}

\newcommand{\NormalDist}[2]{\ensuremath{\mathcal{N}(x,y)}}

\providecommand{\sep}{;}
\providecommand{\keywords}[1]{\noindent \textbf{Keywords:} #1 }
\providecommand{\pacs}[1]{\noindent \textbf{PACS numbers:} #1 }

\begin{document}

\title{Employing online quantum random number generators for generating truly
random quantum states in \Mathematica}

\author{Jaros{\l}aw Adam Miszczak\\
\texttt{miszczak@iitis.pl}\\
Institute of Theoretical and Applied Informatics,\\
Polish Academy of Sciences, Ba{\l}tycka 5, 44-100 Gliwice, Poland
}

\maketitle

\begin{abstract}

We present a new version of \TRQS\ package for \Mathematica\ computing system.
The package allows harnessing quantum random number generators (QRNG) for
investigating the statistical properties of quantum states. It implements a
number of functions for generating random states. The new version of the package
adds the ability to use the on-line quantum random number generator service and
implements new functions for retrieving lists of random numbers. Thanks to the
introduced improvements, the new version provides faster access to high-quality
sources of random numbers and can be used in simulations requiring large amount
of random data.\\


\end{abstract}

\section*{New Version Program Summary}
\noindent\emph{Program title:} TRQS\\
\emph{Program author:} Jaros{\l}aw A. Miszczak\\
\emph{Program web page:} \texttt{http://www.iitis.pl/~miszczak/trqs}\\
\emph{Distribution format:} tar.gz\\
\emph{No. of bytes in distributed program, including test data, etc.:} $218\ 618$\\
\emph{No. of lines in distributed program, including test data, etc.:} $16\ 479$\\
\emph{Programming language:} Mathematica, C\\
\emph{Computer:} any supporting Mathematica in version 7 or higher\\
\emph{Operating system:} any platform supporting Mathematica; tested with
GNU/Linux (32 and 64 bit)\\
\emph{RAM:} case-dependent\\
\emph{Classification:} 4.15\\
\emph{Catalogue identifier of previous version:} AEKA\_v1\_0\\
\emph{Reference in CPC:} Comput. Phys. Commun. 183 (2012) 118 \cite{trqs}\\
\emph{Does the new version supersede the previous version?:} Yes.\\
\emph{Nature of problem:} Generation of random density matrices and utilization
of high-quality random numbers for the purpose of computer simulation.\\
\emph{Method of solution:} Use of a physical quantum random number generator and
an on-line service providing access to the source of true random numbers generated
by quantum real number generator.\\ 
\emph{Running time:} Depends on the used source of randomness and the amount of
random data used in the experiment.\\
\emph{Reasons for new version:} Added support for the high-speed on-line quantum
random number generator and improved methods for retrieving lists of random
numbers.

\section{Summary of revisions} 
The presented version provides two significant improvements. 

The first one is the ability to use on-line Quantum Random Number Generation
service developed by PicoQuant GmbH and the Nano-Optics groups at the Department
of Physics of Humboldt University. The on-line service supported in the version
2.0 of the \TRQS\ package provides faster access to true randomness sources
constructed using the laws of quantum physics. The service is freely available
at \cite{arng}. The use of this service allows using the presented package with
the need of a physical quantum random number generator. 

The second improvement introduced in this version is
the ability to retrieve arrays of random data directly for the used source. This
increases the speed of the random number generation, especially in the case of
an on-line service, where it reduces the time necessary to establish the
connection. Thanks to the speed improvement of the presented version, the package
can now be used in simulations requiring larger amount of random data. Moreover,
the functions for generating random numbers provided by current version of the
package more closely follow the pattern of functions for generating pseudo-random
numbers provided in \Mathematica.\\ 

\section{Speed comparison}
\begin{figure}[tb!]
	\centering
	\includegraphics[width=1\textwidth]{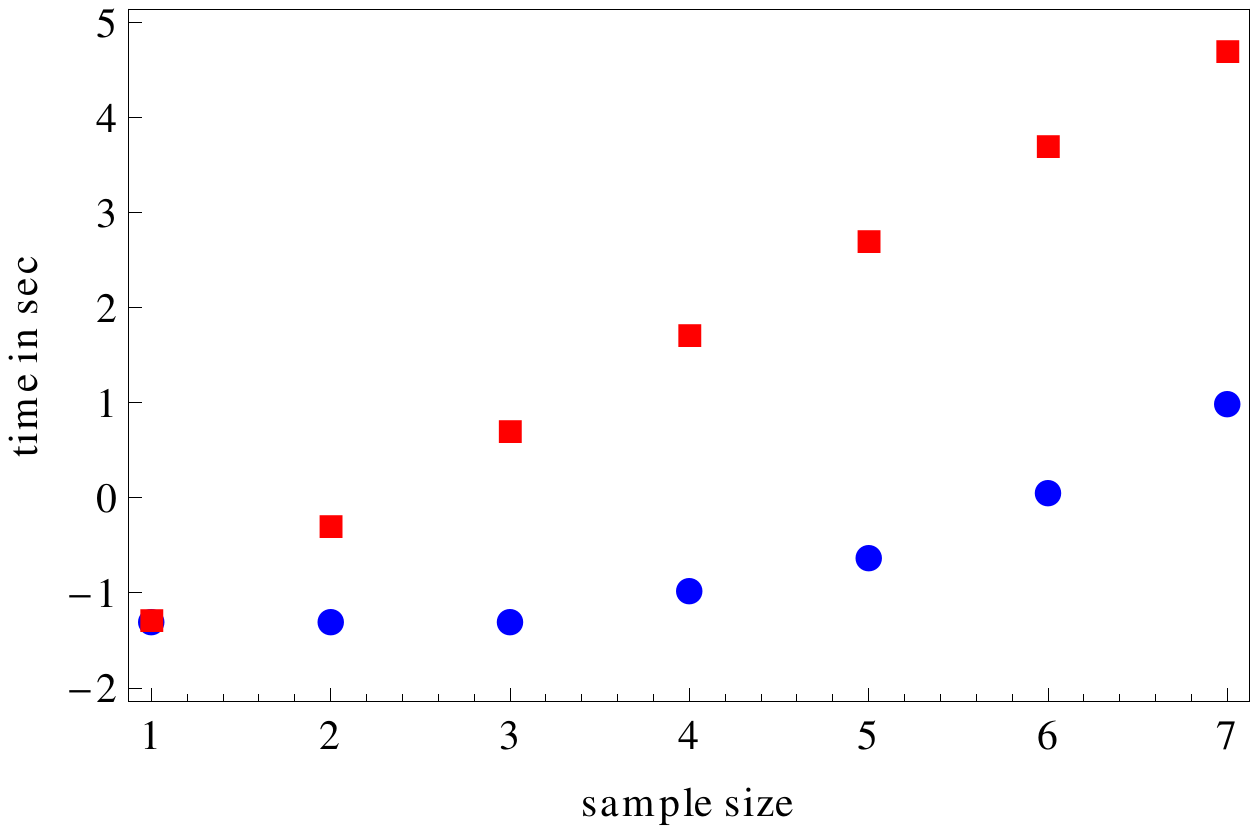}
	\caption{Comparison of the times required to generate a sample of random numbers
	using Quantis device (red squares) and QRNG service (blue dots) in log-log
	scale. In the case of the QRNG service the plot presents time averaged over
	three experiments. In both cases the results were obtained by generating
	matrices of random real numbers from $[0,1]$ using
	\lstinline{TrueRandomReal} function.}
    \label{fig:timings-qrgn-vs-quantis}
\end{figure}

The implementation of the support for QRNG on-line service provides a noticeable
improvement in the speed of random number generation. For the samples of real
numbers of size $10^1, 10^2,\dots, 10^7$ the times required to generate these
samples using Quantis USB device and QRNG service are compared in
Fig.~\ref{fig:timings-qrgn-vs-quantis}.
The presented results show that the use of on-line service provides faster access to
random numbers. One should note, however, that the speed gain can increase or
decrease depending on the connection speed between the computer and the server
providing random numbers.

\section*{Acknowledgements} Author would like to thank R.~Heinen and M.~Wahl for
bringing his attention to Ref.~\cite{wahl11ultrafast} and motivating this work
and to C.~Phillips and helping during the development of the package. This work
was supported by the Polish Ministry of Science and Higher Education under the
grant number IP2011 036371.


\begin{thebibliography}{1}
    \bibitem{trqs} J.A. Miszczak, \emph{Generating and using truly random
        quantum states in Mathematica}, Computer Physics Communications, Vol.
        183, No. 1 (2012), pp. 118-124. arXiv:1102.4598
        DOI:10.1016/j.cpc.2011.08.002
    \bibitem{qrng} QRNG Service available on-line at
        \texttt{https://qrng.physik.hu-berlin.de/}. 
    \bibitem{wahl11ultrafast} M. Wahl, M. Leifgen, M. Berlin, T. R\"ohlicke,
        H.-J. Rahn, O. Benson O., \emph{An ultrafast quantum random number
        generator with provably bounded output bias based on photon arrival time
        measurements}, Applied Physics Letters, Vol. 098, 171105 (2011).
        DOI:10.1063/1.3578456 
\end{thebibliography}
\end{document}